\documentclass{article}
\usepackage{graphicx}
\usepackage{booktabs} 
\usepackage{amsmath}
\usepackage{amsfonts}
\usepackage{authblk}
\usepackage{color}
\usepackage{hyperref}
\usepackage{array}
\usepackage{multirow} 
\usepackage{arydshln}
\usepackage[numbers,sort&compress]{natbib}

\newcommand{\tabref}[1]{(Table~\ref{#1})}

\newcommand{\figref}[2]{(Fig.~\ref{#1}#2)}

\newcommand{\ie}{\textit{i.e.,~}}
\newcommand{\eg}{\textit{e.g.,~}}
\setcitestyle{numbers,square,super,comma,compress}

\usepackage[accepted]{icml2021}

\icmltitlerunning{A Hitchhiker's Guide to Deep Chemical Language Processing for Bioactivity Prediction}

\begin{document}
\twocolumn[
\icmltitle{A Hitchhiker's Guide to Deep Chemical Language Processing for Bioactivity Prediction}

\begin{icmlauthorlist}
\icmlauthor{Rıza Özçelik}{tue,clt}
\icmlauthor{Francesca Grisoni}{tue,clt}
\end{icmlauthorlist}

\icmlaffiliation{tue}{Institute for Complex Molecular Systems and Dept. Biomedical Engineering, Eindhoven University of Technology, Eindhoven, Netherlands.}
\icmlaffiliation{clt}{Centre for Living Technologies, Alliance TU/e, WUR, UU, UMC Utrecht, Netherlands.}

\icmlcorrespondingauthor{Francesca Grisoni}{f.grisoni@tue.nl}

\vskip 0.3in
]

\printAffiliationsAndNotice{}  
\begin{abstract}
\noindent Deep learning has significantly accelerated drug discovery, with `chemical language' processing (CLP) emerging as a prominent approach. CLP learns from molecular string representations (\eg Simplified Molecular Input Line Entry Systems {[SMILES]} and Self-Referencing Embedded Strings [{SELFIES]}) with methods akin to natural language processing. Despite their growing importance, training predictive CLP models is far from trivial, as it involves many `bells and whistles'. Here, we analyze the key elements of CLP training, to provide guidelines for newcomers and experts alike. Our study spans three neural network architectures, two string representations, three embedding strategies, across ten bioactivity datasets, for both classification and regression purposes. This `hitchhiker's guide' not only underscores the importance of certain methodological choices, but it also equips researchers with practical recommendations on ideal choices, \eg in terms of neural network architectures, molecular representations, and hyperparameter optimization.  
\end{abstract}
\section{Introduction}
Machine learning has accelerated drug discovery \cite{vamathevan2019applications,ozccelik2023structure}. The prediction of biological properties, such as the interaction with macromolecular targets, has been pivotal in this context, \eg for hit finding and lead optimization \cite{chakraborty2023utilizing,stokes2020deep, ozccelik2023structure, van2024deep}. Deep learning models that use string representations of molecules, like Simplified Molecular Input Line Entry System (SMILES) \cite{weininger1988smiles} and Self-Referencing Embedded Strings (SELFIES) \cite{krenn2020self}, have drawn particular interest \cite{ozturk2018deepdta,zhao2022attentiondta,bjerrum2017smiles}. Such deep `chemical language' processing approaches apply methods akin to natural language processing to learn from molecular string representations \cite{ozturk2020exploring,ross2022large}.

\begin{figure*}[t]
    \centering
    \includegraphics[width=0.75\textwidth]{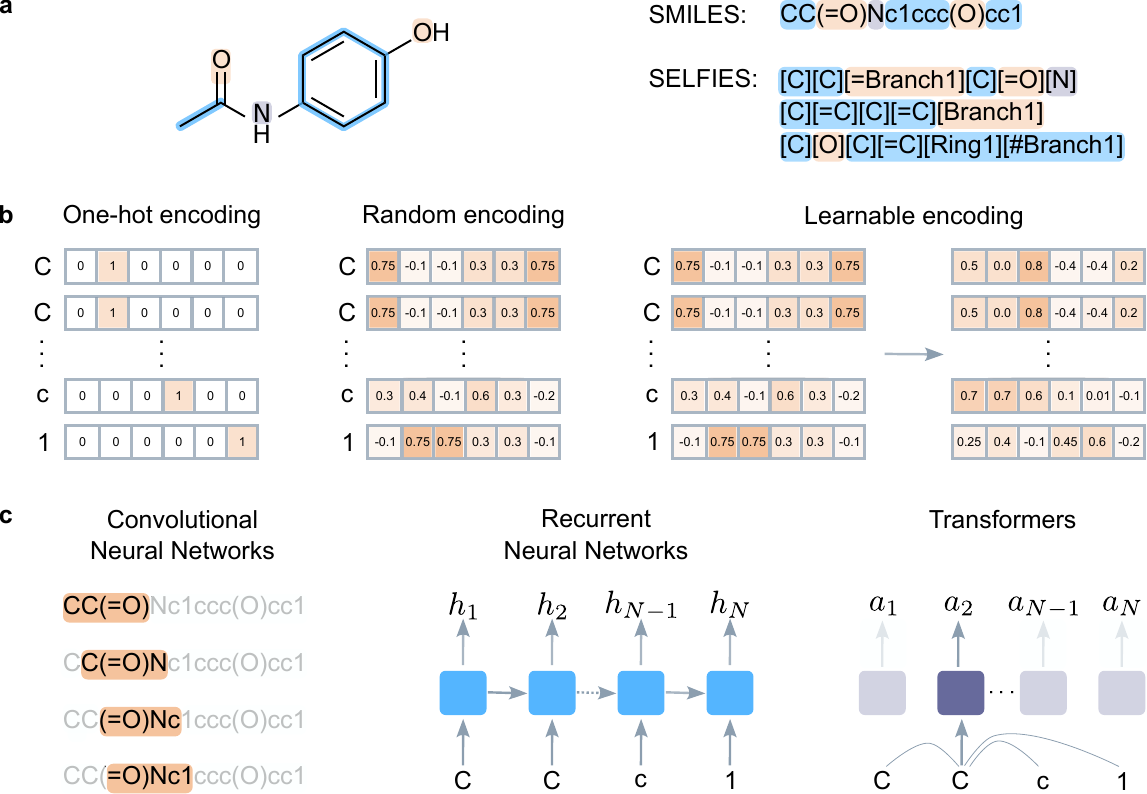}
    \caption{\textit{Deep Chemical Language Processing for Bioactivity Prediction.} \textbf{(a)} String notations such as SMILES and SELFIES represent a molecular graph as a sequence of characters (`tokens'). The atoms are represented with periodic table symbols, while branches, rings, and bonds are assigned special characters. \textbf{(b)} Token encoding, where the chosen molecular string is converted into a matrix to train deep learning models. One-hot encoding represents each token with a unique binary vector. Random encoding maps tokens to fixed, unique, and continuous vectors. Learnable encoding starts with a random vector per token and updates the vectors during training to improve the model performance. \textbf{(c)} Architectures used in this study. Convolutional neural networks slide windows over the input sequences, and learn to weight and aggregate the input elements. Recurrent neural networks iterate over the input tokens in a step-wise manner, and update the `hidden' information learned from the sequence ($h_i$). Transformers learn all-pair relationships between the input tokens and learn to weight each input representation to create the representations in the next layers ($a_i$).}
    \label{fig:figure1}
\end{figure*}

Molecular string representations (\eg SMILES \cite{weininger1988smiles} and SELFIES \cite{krenn2020self}, among others\cite{o2018deepsmiles,wu2024tsmiles,heller2015inchi,noutahi2024gotta}) have found widespread application in cheminformatics and related fields \cite{grisoni2023chemical,ozturk2020exploring}. They convert two-dimensional molecular information into strings, by traversing the molecular graph and annotating atom and bond information with dedicated letters \figref{fig:figure1}{a}. Deep `chemical language processing' (CLP) models are then trained to map the chemical information in such strings to a property to be predicted, \eg a ligand interaction with a target or toxicological properties. Once trained, CLP models can be applied prospectively, for instance, to screen large molecular libraries in search of molecules with desirable properties \cite{kimber2021maxsmi,moret2023leveraging}.

Developing predictive CLP models is far from trivial \cite{van2022exposing,zhou2018exploring} and it requires many choices to be made \cite{bengio2012practical}, \eg in terms of molecular string representations and their encoding, and of neural network architectures and their hyperparameters. Each such choice might affect the model performance. Stemming from these observations, this `hitchhiker's guide' aims to discover best practices in the field, and provide a guideline for what choices to make when training CLP models for bioactivity prediction. Here, we derive our insights from a systematic analysis of three deep learning architectures, two molecular string representations, and three encoding approaches on ten datasets spanning regression and classification tasks. 

Ultimately, this `hitchhiker's guide' provides some `tricks of the trade' and practical recommendations -- for beginners and experts alike -- on what choices to prioritize when training deep chemical language processing models from scratch. We hope that this paper will accelerate the adoption of deep chemical language processing approaches, and spark novel research to further their potential. 

\section{Methods}

\subsection{Molecular String Representations}
String representations capture two-dimensional molecular information as a sequence of characters (`tokens'). Here, we focus on the two most popular string representations \figref{fig:figure1}{a,b}:
\begin{itemize}
    \item \textit{Simplified Molecule Input Line Entry Systems (SMILES)}\cite{weininger1988smiles} strings, which start from any non-hydrogen atom in the molecule and traverse the molecular graph. Atoms are annotated as their element symbols, bonds (except for single bonds) are annotated with special tokens (\eg, `=': double, `$\#$': triple), and branching is indicated by bracket opening and closure. Stereochemical information can also be indicated by dedicated tokens, although this will be not considered in this study. Initially proposed for chemical information storage, SMILES strings constitute, to date, the \textit{de facto} notation in chemical language processing \cite{ozccelik2021chemboost,sharma2021smiles,wu2021learning,kimber2021maxsmi,grisoni2023chemical}.

    \item \textit{Self-Referencing Embedded Strings (SELFIES)}\cite{krenn2020self}, which were recently proposed as SMILES alternatives. SELFIES encode the atoms with their symbols, and annotate their connectivity via branch length, ring size, and referencing previous elements. SELFIES strings have been developed for \textit{de novo} design \cite{krenn2020self,nigam2021beyond,choi2023rebadd,krenn2022selfies} and are finding increasing applications for bioactivity prediction \cite{yuksel2023selformer,feng2024gcardti}. 
\end{itemize}

\subsection{Token Encoding}
For deep learning purposes, molecular strings are converted into sequences of vectors, by `vectorizing' each token in the string. Here, we experimented with three encoding approaches \figref{fig:figure1}{b}, namely: 

\begin{itemize}
    \item \textit{One-hot encoding}, which represents tokens with $V$-dimensional binary vectors,  $V$ being the number of unique tokens (`vocabulary' size). Each token is allocated a different dimension in this space and has a vector on which only that dimension is set to 1, and the rest is set to 0. One-hot encoding ensures that all token vectors are orthogonal to each other, \textit{i.e.}, the similarity between all tokens is zero.
    
    \item \textit{Learnable embeddings}, whereby a random continuous vector is assigned to each token. These vectors are updated (`learned') during training to optimize the predictions. The updates might enable models to learn relationships between parts of the molecules (and the corresponding tokens) that can be useful for bioactivity prediction. 
    
    \item \textit{Random Encoding}, which assigns a continuous vector to each token and uses the same vector throughout the model training. This approach is intermediate between learnable embeddings and one-hot encoding. Like learnable embeddings, the vectors have continuous values, and they are fixed during training like one-hot encoding. 
\end{itemize}

\subsection{Deep Learning Architectures}
We experimented with three well-established deep learning architectures \figref{fig:figure1}{c}. They differ in how they process and combine information on the (encoded) input molecular strings to predict bioactivity. 
\begin{table}
\caption{\ \textit{Datasets used in this study.} We curated ten bioactivity datasets, for classification (\ie binding vs non-binding\cite{sun2017excape}) and regression (\ie $pK_i$ prediction\cite{van2022exposing}) purposes. For each dataset, we report ID, target name, and total number of molecules (\textit{n}).}
\label{tab:datasets}
\resizebox{\linewidth}{!}{ %
\begin{tabular}{llp{4.8cm}c}
\hline
\textbf{Task}           & \textbf{ID}  & \textbf{Target name}         & \textbf{\textit{n}}   \\ \hline
Class. & DRD3(c)& Dopamine Receptor D3 & 5500\\
 & FEN1& Flap Structure-specific Endonuclease 1 & 5500 \\
 & MAP4K2&  Mitogen-activated protein 4x Kinase 2&5500 \\
 & PIN1& Peptidyl-prolyl cis/trans Isomerase & 5500 \\
 & VDR& Vitamin D Receptor & 5500 \\   \hline
 Reg. & MOR& $\mu$-opioid Receptor & 2838 \\
 & DRD3(r)& Dopamine Receptor D3 & 3596 \\
 & SOR & Sigma Opioid Receptor & 1325 \\
 & PIM1& Serine/threonine-protein Kinase PIM1 & 1453 \\ \hline

\end{tabular}
}
\end{table}

\begin{itemize}
    \item \textit{Convolutional neural networks} (CNNs) \cite{lecun1998gradient}. CNNs slide windows (called kernels) over an input sequence, and learn to weight input elements at each window. Such window sliding enables CNNs to capture local patterns in sequences, which are then stacked to predict the global properties of a string (\eg bioactivity). 
    
    \item \textit{Recurrent neural networks (RNNs) \cite{hopfield1982neural}}. RNNs are recurrent models, \ie they iterate over the input token and, at each step, compress the information into a `hidden state'. Here, we used bidirectional RNNs -- which iterate over the sequence in both directions and concatenate the final hidden states to encode the sequence -- in combination with gated recurrent units \cite{cho-etal-2014-learning}.
    
    \item \textit{Transformers} \cite{vaswani2017attention}, which learns patterns between pairs of input tokens, using a mechanism called `self-attention'. Self-attention learns to represent input sequences by learning to weight the link between every token pair. Since self-attention makes transformers invariant to the token position in the sequence, here we adopted learnable positional embeddings to capture the sequence structure.
\end{itemize}

\subsection{Bioactivity Datasets}
We curated ten bioactivity datasets containing 1453 to 5500 molecules \tabref{tab:datasets}, and spanning two tasks, namely (a) classification (5 datasets), \ie predicting whether a molecule is active or inactive on a given target (in the form of a label), and (b) regression (5 datasets), where the coefficient of inhibition ($K_i$) is to be predicted.

\begin{itemize}
    \item \textit{Classification datasets.} Five datasets were curated from ExCAPE-DB \cite{sun2017excape}, which collects ligand-target bioactivity information (in the form of `active'/`inactive') on 1677 proteins. In this work, we selected five targets: dopamine receptor D3 (DRD3), Flap structure-specific endonuclease 1 (FEN1), Mitogen-activated protein kinase kinase kinase kinase 2 (MAP4K2), peptidyl-prolyl cis/trans isomerase (PIN1), and vitamin D receptor (VDR). For each macromolecular target, a set of 5500 molecules (with 10$\%$ of actives) were selected (\textit{see} Section \ref{sec:exp-setup}).
    \item \textit{Regression.} We selected five bioactivity datasets from MoleculeACE \cite{van2022exposing}, which is based on ChEMBL \cite{gaulton2017chembl}. The following datasets were used for $pK_i$ prediction: Serotonin 1a receptor (5-HT1A), $\mu$-opioid Receptor 1 (MOR),  dopamine receptor D3 (DRD3), sigma Opioid Receptor 1 (SOR), and Serine/threonine-protein Kinase PIM1 (PIM1). These datasets were selected to span several target families and to ensure a sufficient number of molecules available for training and testing (from 1453 to 2596).
\end{itemize}

The classification datasets have more molecules than the regression datasets and were built to contain structurally diverse molecules (\textit{see} Section \ref{sec:exp-setup}). Hence, they can be seen as a proxy for hit discovery campaigns, where structurally novel, and bioactive molecules are searched for. Conversely, the regression datasets, which originate from ChEMBL, mostly contain series of highly similar molecules, hence resembling a lead optimization campaign.

\subsection{Performance Evaluation}
The performance of \textit{classification} models was evaluated via the balanced accuracy (BA), expressed as follows: 
\begin{equation}
    \text{BA} = \frac{1}{2}  \left( \frac{TP}{nP} + \frac{TN}{nN} \right),
\end{equation}
\noindent where $TP$ and $TN$ are the numbers of correctly classified positives and negatives, while $nP$ and $nN$ are the total number of positive and negative molecules, respectively.

The performance of \textit{regression} models was evaluated via concordance index \cite{gonen2005concordance,pahikkala2014toward}, which quantifies the model's ability to rank molecules by their experimental potency based on the predicted potency. Both metrics are bound between 0 and 1 -- the closer to 1, the better the performance.

\subsection{Experimental Setup}
\label{sec:exp-setup}
\subsubsection{Data Preparation}
\begin{itemize}
    \item \textit{Classification.} For each selected target, we constructed two sets: (i) Set 1 -- built by randomly sampling 350 actives and 3500 inactives, and (ii) Set 2 -- built by randomly selecting 150 actives and 1500 inactives that were sufficiently distant from Set 1 (\textit{i.e.}, having a minimum edit distance on canonical SMILES strings larger than 10, and a maximum Tanimoto similarity on extended connectivity fingerprints\cite{rogers2010extended} smaller than 60\%). Set 1 was used as a training set, while the molecules of Set 2 were equally divided into a validation and a test set.  
    \textit{\item Regression.} For each target, we created five folds of training validation and test sets (70\%, 15\%, 15\%, respectively). We heuristically minimized train-test similarity by first grouping molecules based on substructure similarity, and then dividing them into training and test set (via \texttt{deepchem}, \texttt{FingerprintSplitter} \cite{Ramsundar-et-al-2019}). 
\end{itemize}

For all collected molecules, we removed stereochemistry, sanitized the molecules, and canonicalized the SMILES strings (\texttt{rdkit v2020.09.01}). We filtered out the molecules with canonical SMILES strings longer than 75 tokens and created the SELFIES strings for all retained molecules. Our data curation pipeline led to different distributions of molecular similarities between training and test set molecules for classification \figref{fig:models}{a} and regression datasets \figref{fig:models}{b}.

\subsubsection{Model Training and Optimization}

We tested all combinations of (a) model architectures (CNN, RNN, and Transformers), (b) molecular strings (SMILES and SELFIES), and encoding approaches (one-hot, random, and learnable) for all datasets. We optimized hyperparameters for each combination and each dataset separately \tabref{tab:hps}. A three-layer perceptron was used as a prediction module for consistency. Finally, XGBoost models\cite{chen2016xgboost} were trained on extended connectivity fingerprints\cite{rogers2010extended} as baselines across all datasets. Early stopping with a patience of five epochs (or trees for XGBoost) and a tolerance of 10$^{-5}$ on validation loss were used. For classification models, we used loss re-weighting to tackle the data imbalance, which assigns the inverted frequency of classes as weights to molecules during loss computation. Finally, the best models were selected based on validation loss, \ie cross-entropy and mean squared error for classification and regression, respectively. 

\begin{table}
    \caption{\ \textit{Model hyperparameters}. Grid search is used to optimize model hyperparameters. Learning rate of 10$^{-2}$ is used only for RNN to balance the number of experiments per architecture.}
    \label{tab:hps}
    \resizebox{\linewidth}{!}{
    \begin{tabular}{lll}
    \hline
    \textbf{Model} & \textbf{Hyperparameter} & \textbf{Search Space} \\
    \hline
    
    All & No. layers & 1, 2, 3 \\
     & Dropout &  0.25 \\
     & Batch size &  32 \\ \hline
    
    CNN & No. filters & 32, 64, 128 \\
     & Kernel length & 3, 5, 7 \\ 
     & Learning rate &  10$^{-2}$, 10$^{-3}$, 5$\times$10$^{-3}$,\\  &  &  10$^{-4}$, 5$\times$10$^{-5}$  \\\hline
    
    RNN & Hidden state dim. & 16, 32, 64, 128 \\
      & Learning rate &  10$^{-2}$
      \\ \hline
    
    Transformer & No. heads & 1, 2, 4 \\
     & MLP dim. & 32, 64, 128 \\
       & Learning rate &  10$^{-2}$, 10$^{-3}$, 5$\times$10$^{-3}$,
       \\  &  &  10$^{-4}$, 5$\times$10$^{-5}$ \\
       \hline

     XGBoost & No. trees & 2000 \\
     (baseline) & Max. depth & 3, 4, 5 \\
     & Eta & 0.01, 0.05, 0.1, 0.2  \\
     & Column fraction &  0.5, 0.75, 1.0 \\
     & Sample fraction & 0.5, 0.75, 1.0 \\ \hline
  \end{tabular}
  }
\end{table}
 
\section{Results}

\begin{figure*}[!t]
    \centering
    \includegraphics[width=0.9\linewidth]{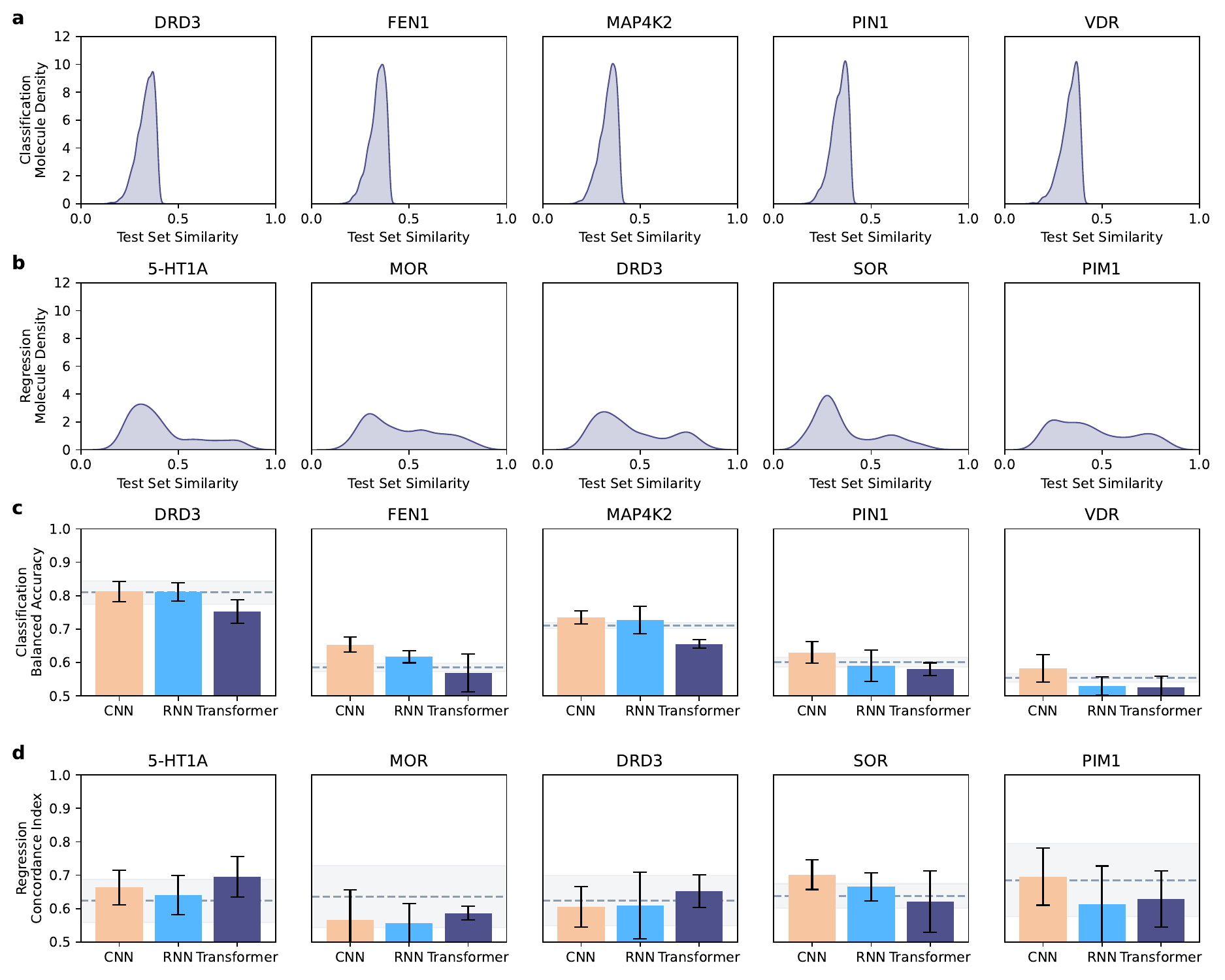}
    
    \caption{\textit{Overview of dataset similarity and of model performance.} \textbf{(a,b)} Distribution of test set similarities in comparison with training set molecules. The similarity was quantified as the Tanimoto coefficient on extended connectivity fingerprints\cite{rogers2010extended}, and the maximum similarity was reported. Different distributions can be observed in the classification (a) and regression (b) datasets, with the former containing more dissimilar molecules on average. \textbf{(c,d)} Performance of neural network architectures across datasets. Bar plots indicate the mean test set performance (with error bars denoting the standard deviation), in comparison with the XGBoost baseline (dashed line: average performance, shaded area: standard deviation). Performance was quantified as balanced accuracy in classification (c), and as concordance index in regression (d).}
    \label{fig:models}
\end{figure*}
\begin{figure*}[!t]
    \centering
    \includegraphics[width=0.9\linewidth]{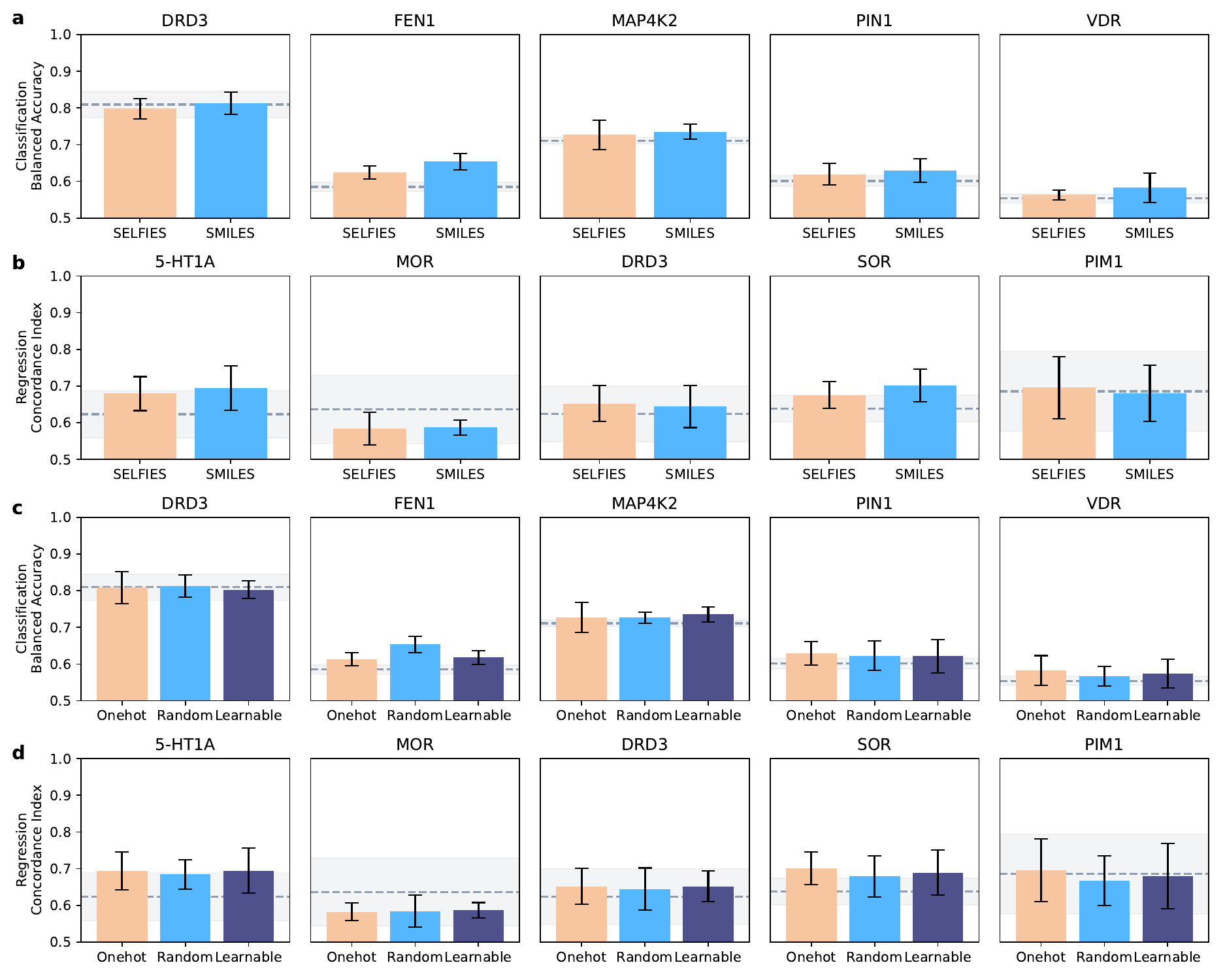}
    
    \caption{\textit{Effect of input molecular strings and of token encoding strategies}. (a,b) Performance of SMILES and SELFIES representations on the model performance. Classification (a) and regression dataset (b) are analyzed separately. (c,d) Performance of token encoding strategies on classification (c) and regression (d). For all plots, bars indicate the mean performance on the test set of each notation, and error bars indicate the standard deviation. The performance of the XGBoost baseline is also indicated (dashed line: average; shaded area: standard deviation).}   
    
    \label{fig:reprs}
\end{figure*}
\subsection{Choosing a Neural Network Architecture}

Here, we aim to gather insights into the effect of the model architecture (CNN \textit{vs} RNN \textit{vs} Transformers) on the performance. To this end, we analyzed the best models per architecture (chosen on the validation set, and analyzed on the test set), regardless of the molecule representation and encoding strategies \figref{fig:models}{c,d}. 

CNNs were consistently the best-performing approach in classification. In regression, CNNs outperform the other approaches on two out of five targets. Transformers yielded the top-performing models in regression (three out of five datasets), while RNNs never yielded the best performance. A Wilcoxon signed-ranked test ($\alpha = 0.05$) on pooled scores across targets per task indicated that CNNs outperform both transformers and RNNs in classification, and RNNs in regression. No statistical differences were observed between CNNs and Transformers in regression.

Interestingly, CNN outperformed the XGBoost baseline in four of five classification datasets, where the test set molecules are structurally dissimilar to the training set (Tanimoto similarity on extended connectivity fingerprints lower than 0.5, Fig, \ref{fig:models}a). In regression, where the test set molecules are more similar to the training set \figref{fig:models}{b}, neither deep models nor XGBoost are statistically superior across the datasets. These results suggest that CLP approaches, and in particular, CNNs might have a higher potential than `traditional' machine learning models when applied to molecules that are structurally diverse from the training set. 

Hence, when considering their performance, architectural simplicity (compared to transformers) and training speed (compared to RNNs), convolutional neural networks constitute the ideal starting choice for chemical language processing and bioactivity prediction. 

\subsection{Representing and Encoding Molecular Structures}

\begin{figure*}[t]
    \centering
    \includegraphics[width=1\linewidth]{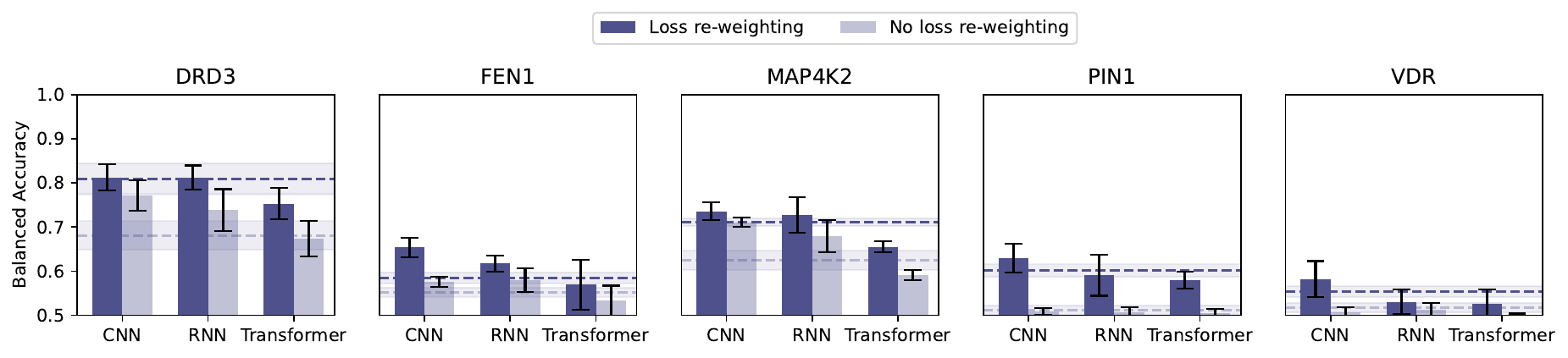}
    
    \caption{\textit{Effect of loss re-weighting.} Comparison of the classification performance obtained with and without loss re-weighting (\ie assigning different weights to the molecules, as the inverse of their class frequency).}
    \label{fig:loss_rw}
\end{figure*}

Here, we aimed to unveil the effect of the chosen molecular string representation (SMILES \textit{vs} SELFIES) and token embedding (one-hot, random, and learnable) strategies. To this end, we compared the best models for each molecule representation and token encoding (minimum average error on the validation set). When investigating for practical guidelines, the differences are less evident than when choosing a neural network architecture \figref{fig:reprs}{}. 

SMILES strings yield higher performance than SELFIES across classification tasks (p < 0.05, Wilcoxon signed-ranked test). In regression, SELFIES outperform SMILES strings on two datasets (DRD3 and PIM1), and show similar performance otherwise, without statistically significant differences (Wilcoxon signed-ranked test, $\alpha = 0.05$). In general, the performance differences due to the chosen string notation are lower than those caused by the model architecture, with few exceptions.

When analyzing the encoding strategies, no approach consistently outperformed the others \figref{fig:reprs}{c,d}, suggesting  that all encoding approaches impact bioactivity prediction comparably. This underscores that, in the space of our design of experiments, choosing model architecture first, and then molecular string notations, should have higher priority than the encoding strategy. 

When considering these results, we recommend CLP hitchhikers \cite{galaxy_42} to use SMILES strings combined with learnable encoding. SMILES strings are, in fact, ubiquitous in available databases, and numerous tools exist to process them (\eg \texttt{rdkit}). This aspect makes SMILES strings easier to work with, with no loss in performance. Learnable representations are also simple to use, and are implemented in most major deep learning libraries (\eg Pytorch,\cite{paszke2019pytorch}, Tensorflow\cite{tensorflow2015-whitepaper}, and Keras\cite{chollet2015}).

\subsection{Other Tricks of the Trade}

While the previous sections have tackled the most important algorithm-design choices in CLP, there are still many `bells and whistles'\cite{bengio2012practical} involved in obtaining predictive models. In what follows, we will focus on the loss function and hyperparameter optimization -- both aspects impacting the effectiveness of the training process, and, ultimately, the model predictivity. 

\noindent \underline{Loss functions for imbalanced classes}. Class imbalance is common in bioactivity datasets \cite{volkamer2023machine}, since desirable outcomes (\eg bioactive or non-toxic molecules) occur less frequently. Surprisingly, public bioactivity databases might be \textit{unrealistically} imbalanced, with a lack of ‘negative’ data (\eg inactive molecules) due to reporting bias. Hence, mitigating the negative effects of class imbalance on the model performance is key for CLP hitchhikers \cite{galaxy_42}.

To mitigate class imbalance, in all the classification results shown so far, we applied loss re-weighting. We assigned a weight of 10 to the active molecules and of 1 to the inactive molecules (corresponding to the inverse of their respective class frequency). Loss re-weighting substantially increased balanced accuracy of 6\% on average \figref{fig:loss_rw}{}. In some extreme cases (\ie PIM1 and VDR) the lack of loss re-weighting led to a balanced accuracy of 0.5 (baseline-level performance). Class re-weighting is hence a simple and effective strategy that we recommend, among other options\cite{wang2020comprehensive}.
\begin{figure*}[!t]
    \centering
    \includegraphics[width=.59\linewidth]{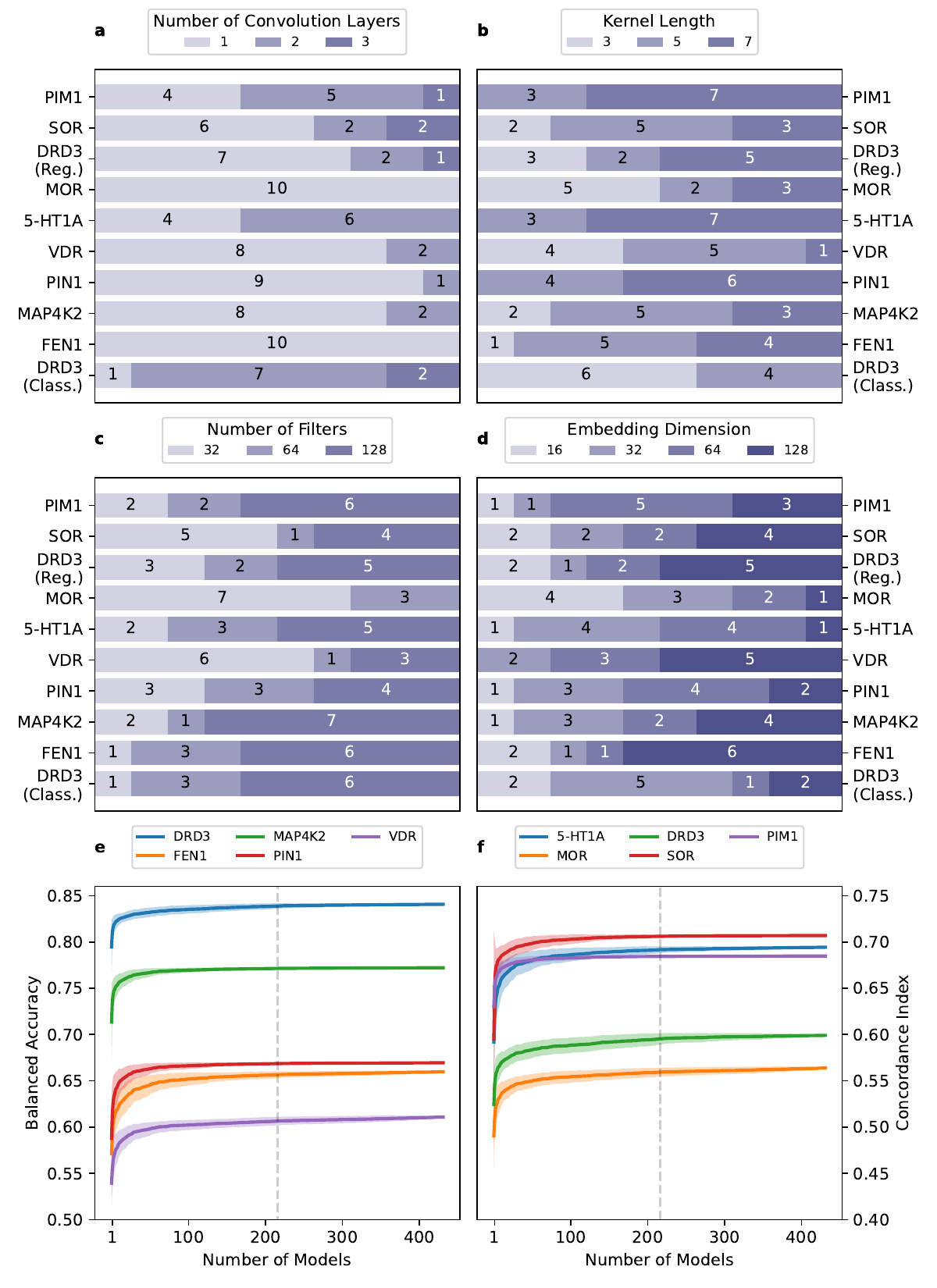}
    
    \caption{\textit{Hyperparameter tuning.} (a-d) Most frequently occurring hyperparameter values among the top-ten models per dataset (CNN architecture, with SMILES strings and learnable embeddings). The following parameters were investigated: number of convolution layers (a), kernel length (b), number of filters (c), and token embedding dimension (d). (e,f) Model performance vs. explored hyperparameter space size. Performance of progressively subsampled models from 1 to 432 hyperparameter configurations (total) for both classification (e) and regression (f). The dashed line indicates 50\% of models being explored.
    }
    
    \label{fig:hps}
\end{figure*}
\noindent \underline{Optimal hyperparameters}. Hyperparameter optimization can be a demanding task due to the high number of hyperparameters to explore and required domain expertise. To equip CLP practitioners with guidelines, we focused on our recommended setup (CNNs trained on SMILES strings with learnable embedding), and inspected the top-10 performing models (the test set average) for the following hyperparameters \figref{fig:hps}{}: (a) \textit{number of convolution layers}, impacting the network depth and complexity, (b) \textit{kernel length}, controlling the size of learned patterns, (c) \textit{number of convolution filters}, controlling information compression across layers, and (d) \textit{token embedding dimension}, controlling the size of the latent representations learned. 

The best-performing models tend to have a low number of layers, with one being the most prevalent (seven out of ten datasets, and 67\% occurrence, Fig. \ref{fig:hps}a). Optimal kernel size and number of filters \figref{fig:hps}{b,c} results are dataset dependent. Finally, embeddings of 32 or higher dimensions are preferred (84\% of cases, Fig. \ref{fig:hps}d). These results offer indications for hyperparameter prioritization `on a budget', although we recommend conducting extensive searches whenever feasible.

\noindent \underline{Exploring the hyperparameter galaxy}. To provide guidelines for parsimonious hyperparameter optimization, we randomly sampled an increasing number of models from the hyperparameter space (from 1 to the total, 432), and analyzed the performance of the top-ten models \figref{fig:hps}{e,f}. Performance often plateaued before reaching 100 models, with a shrink in its variability when half of the space was explored. These findings indicate that defining a high-dimensional hyperparameter space can be better than relying on a narrow one, and that randomly exploring half of the grid can be sufficient to reach the maximum performance level possible in that space.
\section{So Long, and Thanks for All the Data}
Casting molecular tasks as chemical language processing has achieved enormous success in the molecular sciences\cite{grisoni2023chemical,ozturk2020exploring}, owed to a unique combination of simplicity (\eg in representing and processing molecules as strings) and performance \cite{flam2022language,ozturk2016comparative}. The importance of chemical language processing is hence only expected to increase. To accelerate the adoption of CLP approaches by novices and experts alike, these are our guidelines for hitchhikers\cite{galaxy_42}, based on the data we have collected:
\begin{enumerate}
    \item \textit{`KISS: Keep It Simple, Silly!'} Convolutional neural networks -- an architecture that is simpler than the Transformer and faster than Recurrent Neural Networks -- yielded the best performance overall, and are recommended as the first choice. Since representation and encoding strategies minimally affected performance, we recommend using SMILES strings for their ubiquity in databases and software, and learnable embeddings for existing implementations in most deep learning packages.
    
    \item \textit{`Cut your losses'}. Molecular bioactivity datasets are inherently imbalanced \cite{volkamer2023machine}, and the `losses' due to such imbalance should be minimized to ensure predictivity \cite{fernandez2018learning}. We recommend loss re-weighting as a simple and yet effective strategy to increase model performance.  
    
    \item `\textit{Cast a wide fishing net}'. Hyperparameter optimization can be computationally demanding. Here, we show that, in general, networks with a low (one to two) number of layers tend to perform well enough, while other hyperparameter choices depend on the dataset. In general, once a hyperparameter space is defined, optimal hyperparameters are likely to be found by exploring half of the possible combinations. Hence, we recommend casting a broad (rather than a narrow) hyperparameter grid for exploration, and refine the hyperparameter values at a later stage.
\end{enumerate}

Several other fascinating properties of the `chemical language' can further the potential of CLP approaches. One of them is molecular string augmentation\cite{bjerrum2017smiles}, where multiple molecular strings can be used to represent the same molecule, \eg to increase the number of data available for training \cite{li2022novel,kimber2018synergy}, or for uncertainty estimation\cite{kimber2021maxsmi,birolo2024deep}. Moreover, transfer learning\cite{cai2020transfer} can be particularly effective on molecular strings\cite{uludougan2022exploiting,moret2023leveraging}, \eg to mitigate the limited data availability on a specific target. We encourage `CLP hitchhikers' to venture forth into such elements and assess their effectiveness on a case-by-case basis.


\section*{Author contributions}
\textit{Conceptualization}: both authors. \textit{Data curation}: R\"{O}. \textit{Formal Analysis}: both authors. \textit{Investigation}: both authors. \textit{Methodology}: both authors. \textit{Software}: R\"{O}. \textit{Visualization}: R\"{O}. \textit{Writing – original draft}: R\"{O}.
\textit{Writing – review and editing}: both authors.

\section*{Acknowledgements}
This research was co-funded by the European Union (ERC, ReMINDER, 101077879). Views and opinions expressed are however those of the author(s) only and do not necessarily reflect those of the European Union or the European Research Council. Neither the European Union nor the granting authority can be held responsible for them. The authors also acknowledge support from the Irene Curie Fellowship and the Centre for Living Technologies.

\bibliography{references}
\bibliographystyle{ieeetr}

\end{document}